\documentclass{aa}  
\usepackage{graphicx, array}
\usepackage[varg]{txfonts}
\usepackage{natbib,twoopt}
\usepackage{xcolor}
\usepackage[breaklinks=true]{hyperref} 

\newcommand{\kms}{km~s$^{-1}$}
\newcommand\gdrtwo{\gaia~DR2}
\newcommand{\gaia}{\textit{Gaia}}

\begin{document}

   \title{Open cluster kinematics with \gaia\ DR2 
   \thanks{The table with clusters velocities is only available in electronic form at the CDS via anonymous ftp to cdsarc.u-strasbg.fr (130.79.128.5) 
or via http://cdsarc.u-strasbg.fr/viz-bin/qcat?J/A+A/?/?}}

    \author{
C. Soubiran\inst{\ref{LAB}}
\and
T. Cantat-Gaudin\inst{\ref{IEECUB}}
\and
M. Romero-Gomez\inst{\ref{IEECUB}}
\and
L. Casamiquela\inst{\ref{LAB}}
\and
C. Jordi\inst{\ref{IEECUB}}
\and
A. Vallenari\inst{\ref{OAPD}}
\and
T. Antoja\inst{\ref{IEECUB}}    
\and
L. Balaguer-N{\'u}{\~n}ez\inst{\ref{IEECUB}}    
\and
D. Bossini\inst{\ref{OAPD}}
\and
A. Bragaglia\inst{\ref{OABO}}
\and
R. Carrera\inst{\ref{OAPD}}
\and
A. Castro-Ginard\inst{\ref{IEECUB}}
\and
F. Figueras\inst{\ref{IEECUB}}
\and
U. Heiter\inst{\ref{UU}}
\and
D. Katz\inst{\ref{GEPI}}
\and
A. Krone-Martins\inst{\ref{SIMUL}}
\and
J.-F. Le Campion\inst{\ref{LAB}}
\and
A. Moitinho\inst{\ref{SIMUL}}
\and
R. Sordo\inst{\ref{OAPD}}
}

\institute{
Laboratoire d'Astrophysique de Bordeaux, Univ. Bordeaux, CNRS, B18N, all\'ee Geoffroy Saint-Hilaire, 33615 Pessac, France\label{LAB}
\email{caroline.soubiran@u-bordeaux.fr}
\and
Institut de Ci\`encies del Cosmos, Universitat de Barcelona (IEEC-UB), Mart\'i i Franqu\`es 1, E-08028 Barcelona, Spain\label{IEECUB}
\and
INAF-Osservatorio Astronomico di Padova, vicolo Osservatorio 5, 35122 Padova, Italy\label{OAPD}
\and
INAF-Osservatorio di Astrofisica e Scienza dello Spazio, via Gobetti 93/3, 40129 Bologna, Italy\label{OABO}
\and
CENTRA, Faculdade de Ci\^encias, Universidade de Lisboa, Ed. C8, Campo Grande, P-1749-016 Lisboa, Portugal\label{SIMUL}
\and
Department of Physics and Astronomy,  Uppsala University, Box 516, 75120 Uppsala, Sweden\label{UU}
\and
GEPI, Observatoire de Paris, Universit\'e PSL, CNRS, 5 Place Jules Janssen, 92190 Meudon, France \label{GEPI}
}
  

 \date{Received \today, accepted }

 
  \abstract
   {Open clusters are very good tracers of the evolution of the Galactic disc. Thanks to \gaia, their kinematics can be investigated with an unprecedented precision and accuracy.}
   {The distribution of open clusters in the 6D phase space is revisited with \gdrtwo.}
   {The weighted mean radial velocity of open clusters was determined, using the most probable members available from a previous astrometric investigation that also provided mean parallaxes and proper motions.  Those parameters, all derived from \gdrtwo\ only, were combined to provide the 6D phase-space information of 861 clusters. The velocity distribution of nearby clusters was investigated, as well as the spatial and velocity distributions of the whole sample as a function of age. A high-quality subsample was used to investigate some possible pairs and groups of clusters sharing the same Galactic position and velocity. }
 {For the high-quality sample of 406 clusters, the median uncertainty of the weighted mean radial velocity is 0.5 \kms. The accuracy, assessed by comparison to ground-based high-resolution spectroscopy, is better than 1 \kms. Open clusters nicely follow the velocity distribution of field stars in the close solar neighbourhood as previously revealed by \gdrtwo. As expected, the vertical distribution of young clusters is very flat, but the novelty is the high precision to which this can be seen. The dispersion of vertical velocities of young clusters is at the level of  5 \kms.  Clusters older than 1 Gyr span distances to the Galactic plane of up to 1 kpc with a vertical velocity dispersion of 14 \kms, typical of the thin disc. Five pairs of clusters and one group with five members might be physically related. Other binary candidates that have been identified previously are found to be chance alignments.}
   {}

   \keywords{
          stars: kinematics and dynamics -- Galaxy: open clusters -- Galaxy: kinematics and dynamics}

   \maketitle
%

\section{Introduction}
Open clusters (OCs)  are tracers of the formation and evolution of our Galaxy. Their ages cover the entire lifespan of the Galactic disc, tracing the young to old thin-disc components. Their spatial distribution and motion can help to better understand the gravitational potential and the perturbations that act on the structure and dynamics of the Galaxy. Understanding how OCs evolve and disrupt is very important for explaining the assembly and evolution of the Milky Way disc and spiral galaxies in general. Most Galactic OCs evaporate entirely in some $10^8$ years \citep{1971A&A....13..309W}, and the OCs known to be older than 1 Gyr are thought to have survived as a result of their orbital properties, which keep them away from the Galactic plane \citep{1995ARA&A..33..381F}. Internal interactions between members, stellar evolution, encounters with giant molecular clouds, and gravitational harassment by the Galactic potential are the dynamical processes that contribute to the disruption of an OC \citep[see e.g. ][]{2006MNRAS.371..793G, 2016A&A...593A..85G}. The OCs that have survived these effects are thus crucial targets for understanding how several hundreds of thousands of similar objects may already have been dissolved into our Galaxy \citep{2010ApJ...713..166B}. Another important question related to star formation is whether OCs tend to form in pairs or groups. Binary clusters are fairly well established in the Magellanic Clouds, but not in our Galaxy \citep{1995A&A...302...86S, 2009A&A...500L..13D, 2010A&A...511A..38V}. The fraction of binary clusters can shed light on the star-forming activity in molecular clouds and on the tidal disruption timescales. Therefore the determination of the spatial and kinematical properties of OCs and a better knowledge of how they evolve with time provide strong constraints for testing the dynamical processes that occur at local and Galactic scales.

The information about OCs is compiled in large catalogues and databases, such as WEBDA \citep{2003A&A...410..511M}, and the regularly updated catalogues of \cite{2002A&A...389..871D}, hereafter DAML, and of \cite{2013A&A...558A..53K}, hereafter MWSC. With this observational material available before the \gaia\ era, several studies have drawn a picture of the kinematical behaviour of the OC system using several hundred objects. \cite{2005ApJ...629..825D} compiled a sample of 212 clusters for which proper motions, radial velocities, distances, and ages were available, which later became DAML. The authors used the sample to study the pattern speed of the nearby spiral structure. \cite{2009MNRAS.399.2146W} analysed  the kinematics and orbits for a sample of 488 OCs extracted from DAML. They determined the velocity ellipsoid and  computed orbits with three different potentials resulting in different vertical motions. They showed that the distribution of derived orbital eccentricities for OCs is very similar to that derived for the field population of dwarfs and giants in the thin disc.  \cite{2010MNRAS.407.2109V} also analysed DAML and computed the orbits of 481 OCs in order to find clues on their origin. They found that orbital eccentricity and maximum height are correlated with metallicity.  They suggested that four OCs with high altitude and low metallicity could be of extragalactic origin. \cite{2012AstL...38..506G} built a catalogue of fundamental astrophysical parameters for 593 OCs of the Galaxy mainly from the sources mentioned above. The authors compared the kinematical and chemical properties of OCs to those of a large sample of field thin-disc stars and concluded that the properties differ. The authors also found evidence for the heterogeneity of the OC population. \cite{2017A&A...600A.106C} determined 6D phase-space information for 432 OCs and compact associations by combining several catalogues of individual stars and OC parameters \citep{2004AN....325..740K, 2005A&A...438.1163K, 2007AN....328..889K} updated with the mean radial velocity (RV) of 110 OCs determined by \cite{2014A&A...562A..54C} with RAVE data \citep{2006AJ....132.1645S}. They focused on the detection of groups. They identified 19 groupings, including 14 pairs, 4 groups with 3-5 members, and a complex with 15 members. They investigated the age spread and spatial distributions of these structures.

The study of OCs greatly benefits from  \gaia\ data \citep{gdr1a} and in particular from the recent second release, \gdrtwo\  \citep{2018A&A...616A...1G}. \citet[][ hereafter Paper I]{2018arXiv180508726C} determined membership and astrometric parameters for 1\,237 OCs using only \gdrtwo\ data. This very homogeneous sample revealed the distribution of OCs within 4 kpc from the Sun with an unprecedented precision. Furthermore, 60 new OCs were  serendipitously discovered.

\gdrtwo\ provides the RV of about seven million relatively bright, late-type stars \citep{2018A&A...616A...6S} collected by the RVS instrument  \citep{2018A&A...616A...5C}. The combination of parallax, proper motion, and RV gives access to the phase-space information. An illustration of the great potential of \gdrtwo\ for studying the kinematics of the  Galactic disc is given by \cite{2018A&A...616A..11G} and  \cite{2018arXiv180410196A}, who revealed the richness of phase-space substructures. 

The dramatic improvement of the 3D OC velocities with \gdrtwo\ allows us to revisit the global properties of the OC system. In this paper we compute mean RVs of 861 OCs from Paper I using only \gdrtwo\ data. Section \ref{s:RV} describes the procedure we used, the assessment of the precision and accuracy of the catalogue by comparison to ground-based datasets, and the definition of a high-quality sample. We combine the mean RVs with the astrometric parameters derived in Paper I to compute their 3D Galactic velocities (Sect. \ref{s:velocities}). We investigate the kinematics versus age of that sample and try to identify pairs and groups.

\section{Mean radial velocity of open clusters}
\label{s:RV}
\subsection{Input data}
\gdrtwo\ provides the largest and most homogeneous catalogue of RVs for 7.2 million FGK-type stars brighter than G$_{\rm RVS}$ = 12 mag. The stars are distributed over the full celestial sphere. The typical precision of  \gdrtwo\ RVs
is at the \kms\ level. At the bright end, the precision is of the order of 0.2 to 0.3 \kms. At the faint end, it ranges from $\sim$1.4 \kms\ for K stars to $\sim$3.7 \kms\ for F stars.
The \gaia\ spectroscopic processing pipeline is described in  \cite{2018A&A...616A...6S}, while \cite{ 2018arXiv180409372K} describe the properties of the \gaia\ RV catalogue. Additional information useful for the catalogue users can be found in the online documentation \citep{2018gdr2.reptE....V, 2018gdr2.reptE...6S}.

Membership and mean astrometric parameters were investigated in Paper I for 1\,237 OCs. We took the corresponding list of probable members as the starting catalogue for the RV study. Nearly $10\,000$ stars in more than $1\,000$ OCs were found to have an RV in \gaia\ DR2. Their errors range from $0.11$ to $20$ \kms\ with a median value of $1.7$ \kms. The mean RV of an OC, RV$_{\rm OC}$, was computed with a weighting scheme based on the uncertainty of the individual measurements following \cite{sou13}, with an iterative rejection of outliers  differing by more than 10 \kms\ from the mean.
The weight $w_i$ applied to the individual velocity measurement RV$_i$ is $w_i=1/\epsilon_i^2$, $\epsilon_i$ being the RV error for the star $i$ provided in \gdrtwo.  The internal error of  RV$_{\rm OC}$ is $$I=\underset{i}{\sum} w_i\, \epsilon_i /\underset{i}{\sum} w_i$$.
The  RV$_{\rm OC}$ weighted standard deviation $\sigma_{\rm RV_{OC}}$ is defined as 
$$\sigma_{\rm RV_{OC}}^2=\frac{\underset{i}{\sum} w_i}{(\underset{i}{\sum} w_i)^2 - \underset{i}{\sum} w_i^2 }\,
\underset{i}{\sum} w_i ({\rm RV}_i -{\rm {RV_{OC}}})^2$$

The $\rm {RV_{OC}}$ uncertainty is defined as the maximum of the standard error $\sigma_{\rm RV_{OC}} / \sqrt N$ and $I / \sqrt N$  \citep{1988A&A...203..329J},  where  $N$ is the number of star members.

\begin{figure}[ht!]
\begin{center}
\includegraphics[width=\columnwidth]{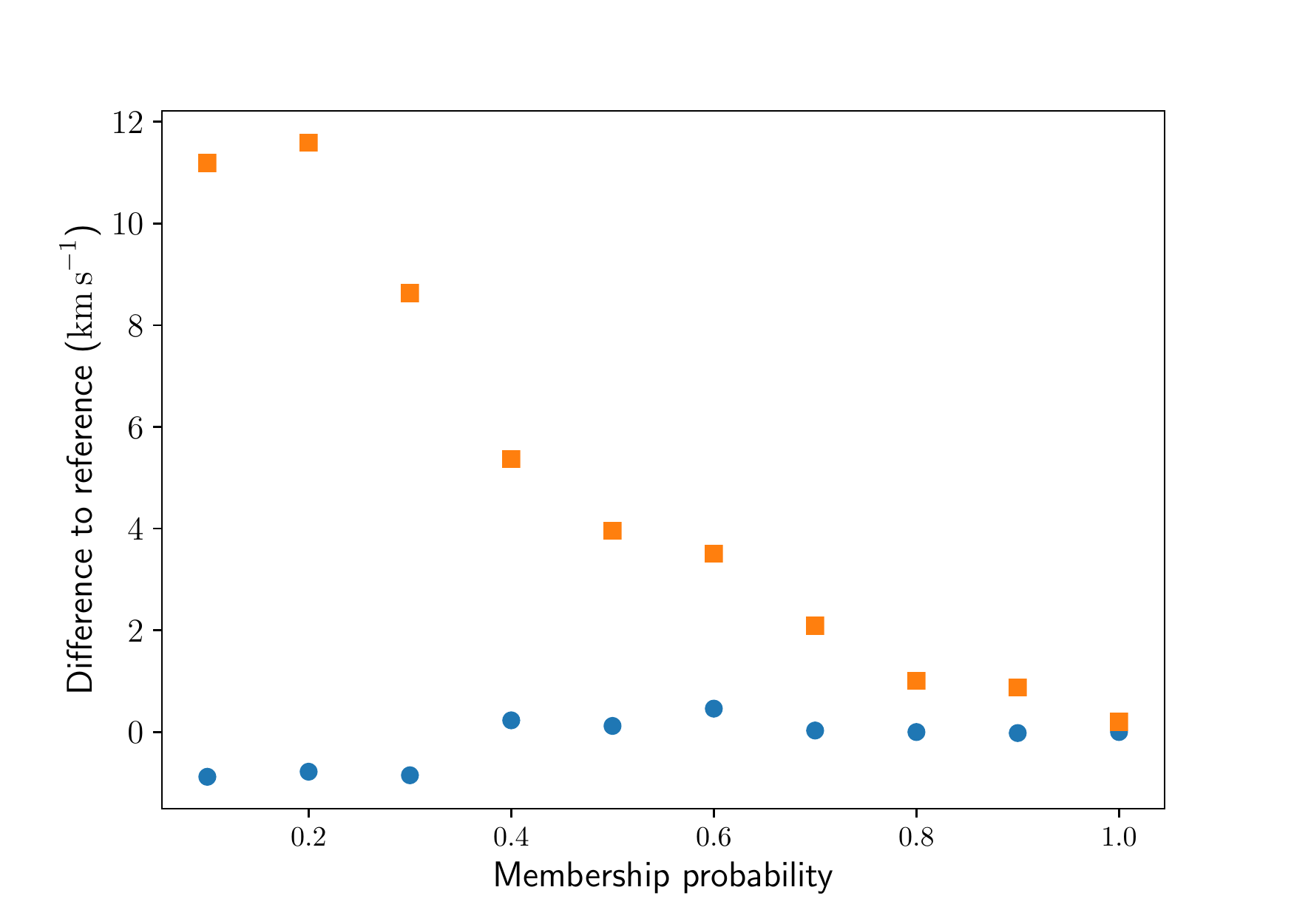}      
\caption{Mean RV for the 312 best OCs computed  as reference value using the most probable members. The median difference to the reference value (blue dots) and MAD (orange squares) is shown for stars with a given probability value.}
\label{f:proba_cut}
\end{center}
\end{figure}

\begin{figure}[ht!]
\begin{center}
\includegraphics[width=\columnwidth]{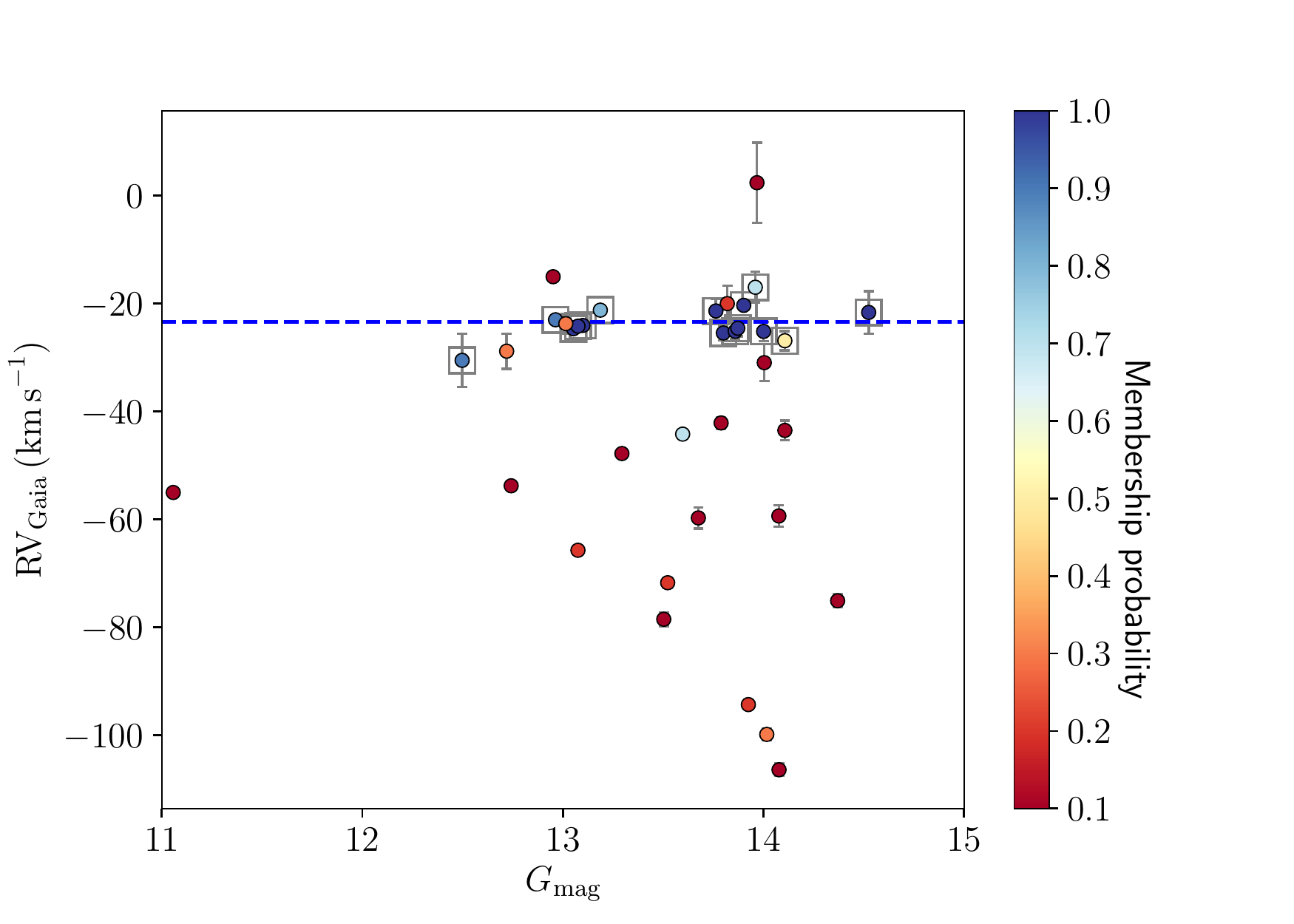}      
\caption{Distribution of RV as a function of $G$ magnitude for the 36 members of Skiff J0058+68.4 with a colour code corresponding to the membership probability, shown as an example. The error bars are the RV uncertainties provided in the \gaia\ DR2 catalogue. The 15 stars used for the calculation of the weighted RV mean are shown as open squares. The mean value, -23.4 \kms, is shown as a blue dotted line.}
\label{f:skiff}
\end{center}
\end{figure}

The list of members in Paper I is provided together with a probability $p$ computed by the UPMASK method \citep{2014A&A...561A..57K}. The probability $p$  can take discrete values of 0.1, 0.2, ... , 0.9, or 1.0.  Of the stars with an RV, $46$\% have $p = 1$ to belong to its parent cluster, while $22$\% have $p < 0.5$. The mean RV of an OC can be significantly different when all the candidate members or only the most probable members are considered. The total number of OCs for which a mean RV can be computed also depends on the adopted probability cut, as shown in Table \ref{t:counts}. It is thus needed to find the optimal selection of stars according to their membership probability that gives the best trade-off between the number of OCs and the uncertainties of RV means. In order to find that optimal cut in membership probability, we compared for a reference subsample the results obtained in each probability class. As reference subsample we selected the 312 OCs with at least 3 members with $p \ge 0.8$ and an uncertainty of the RV mean lower than 2 \kms.  The results obtained for the reference subsample are presented in Fig. \ref{f:proba_cut}, which shows the median difference to the reference value obtained when only stars of a given probability (0.1, 0.2, ..., 0.9, 1.0) are considered. The median absolute deviation (MAD) is also shown. This figure shows that the mean RV of OCs does not change significantly when it is computed with only stars of a given probability, the agreement with the reference value is better than 1 \kms. However, the dispersion increases when less probable members are used to compute the mean RV, owing to the inclusion of non-members with different RVs. This analysis convinced us to consider only stars with a membership probability of $ p \ge 0.4$ because they do not seem highly contaminated by non-members (dispersion $\le 5.5$ \kms).  We find it safer to adopt this rule in order to increase the reliability of the results for OCs with only one member with RV. An illustration of the RV distribution dependence on the membership probability is shown in Fig. \ref{f:skiff} for the OC Skiff J0058+68.4, which has 36 members, 16 with $ p \ge 0.4,$ and one outlier.

\begin{table}[h]
  \centering 
  \caption{Number of stars with a membership probability ($p$) above a given cut, and the corresponding number of OCs with at least one RV measurement. }
  \label{t:counts}
\begin{tabular}{lcrrrrr}
\hline
  cut    & $N$ stars &  N OCs  \\  
\hline
$p \ge 0.1$  & 9\,883 &   1\,039  \\
$p \ge 0.2$ & 8\,983 &  990 \\
$p \ge 0.3$ &  8\,416 &   948 \\
$p \ge 0.4$ & 8\,004 &   907\\
$p \ge 0.5$ &   7\,665 &  873 \\
$p \ge 0.6$ &  7\,291 &  821 \\
$p \ge 0.7$ &  6\,854 &  766 \\
$p \ge 0.8$ & 6\,353 &  713 \\
$p \ge 0.9$ & 5\,672 & 639  \\
$p = 1.0$ & 4\,576 & 531  \\
 \hline
\end{tabular}
\end{table}

\begin{figure}[ht!]
\begin{center}
\includegraphics[width=\columnwidth]{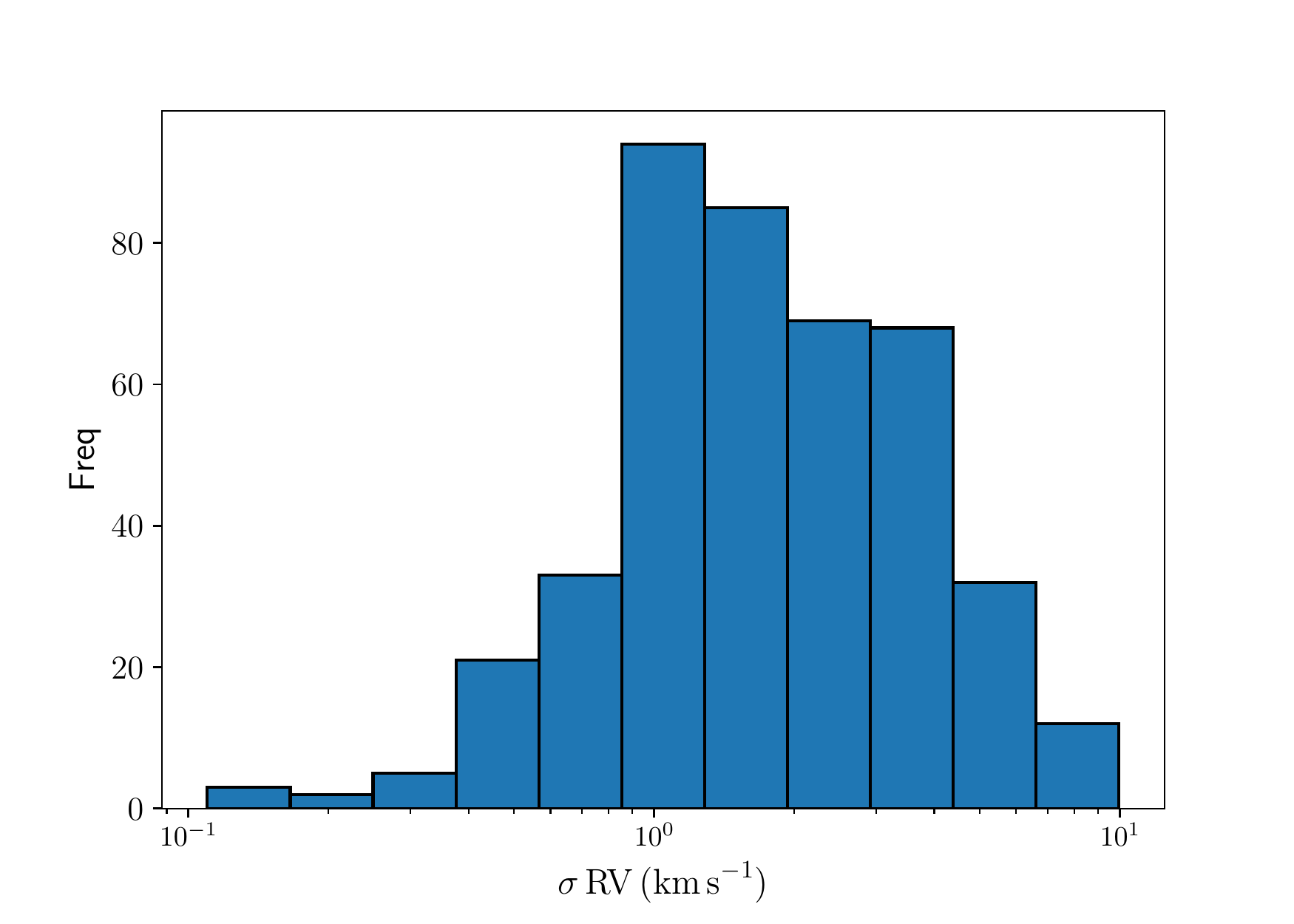}      
\caption{Histogram of the RV standard deviation, in log scale, for the OCs with at least three members. }
\label{f:Rvstd_histo}
\end{center}
\end{figure}

In total we provide a mean RV based on \gaia\ data for 861 OCs. Not all of them have the same level of reliability, 35\% rely on only one star, while nearly 50\% rely on at least three members. The RV standard deviation of this latter subsample is shown in Fig. \ref{f:Rvstd_histo}. The most probable value of the standard deviation lies between 1.0 and 1.5 \kms. The catalogue is available as an electronic table at the CDS.  It gives for each OC the number of members with $p \ge 0.4$, the weighted mean, standard deviation, uncertainty, and the number of members that were kept for the computation after rejection of outliers.

\subsection{Comparison to other datasets}

\begin{figure}[ht!]
\begin{center}
\includegraphics[width=\columnwidth]{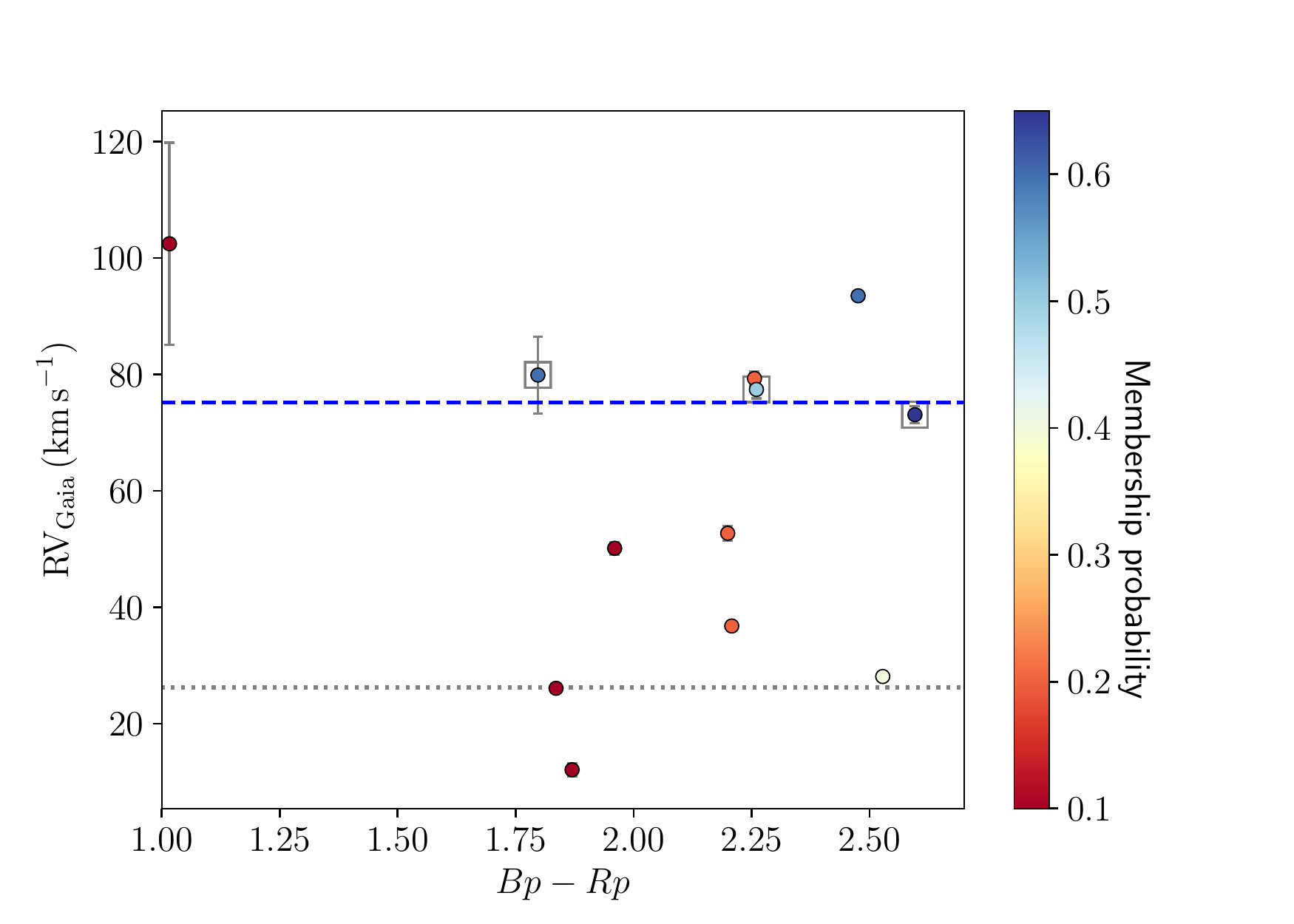}      
\caption{Distribution of RV for members for NGC 2244. The symbols and colour code are the same as in Fig. \ref{f:skiff}. The blue dotted line shows the mean RV computed in this study, while the grey dotted line corresponds to the mean RV quoted in the MWSC.}
\label{f:NGC_2244}
\end{center}
\end{figure}

We compared our mean RV with that of different catalogues. The largest RV source for OC candidates is the MWSC, which provides basic parameters for 3006 entries, including RVs for 953 of them. With this catalogue, we obtain a median difference of 1 \kms\ with a MAD of 5.5 \kms\ for 374 OCs in common. Several outliers with an RV difference larger than 20 \kms\ mostly correspond to OCs in which the RV relies on one or two members, either in the MWSC or in our sample. When restricted to the 140 common OCs with at least three RV members in each catalogue, the median difference remains the same, but the MAD decreases to 2.5 \kms. There are still outliers, such as FSR 0866, for which the \gaia\ RV is $65.5\pm 0.5$ \kms\ based on three members, while the MWSC RV is $0.7\pm20.3$ \kms\ based on two stars,     HIP 33219 and HIP 33287. HIP 33219 (Gaia DR2 888516072258008704) is a B8 star whose parallax is incompatible with the mean parallax of the cluster given in Paper I, while HIP 33287 is a double star that is not in \gdrtwo, but its Hipparcos parallax is also incompatible with the cluster value in Paper I. We conclude that the MWSC RV is based on two non-members. FSR 0866 is a poorly studied OC for which no other RV determination is available in the literature. Another discrepant OC is NGC 2244, for which the \gaia\ RV is $75.2\pm1.8$ \kms\ based on three members (five members with $p \ge 0.4,$ but two outliers), while MWSC gives $26.2\pm3.4$ \kms\  based on 12 hot stars, most of which are astrometric members according to Paper I, but no \gaia\ RV is available for them. The MWSC individual RVs seem of good quality, while the 12 \gaia\ RVs are more dispersed,  as shown in Fig. \ref{f:NGC_2244}. The 3 stars on which the mean \gaia\ RV is based have never been studied before, so that their measurement cannot be assessed against an external source. For this object the MWSC mean RV looks more reliable, although there is no clear explanation why the \gaia\ RV would be in error. According to the Simbad database, NGC 2244 is an ionising cluster of the Rosette Nebula with a nearby associated stellar cluster, NGC 2237. It is therefore possible that two clusters with different RV overlap in that area.

\begin{figure}[ht!]
\begin{center}
\includegraphics[width=\columnwidth]{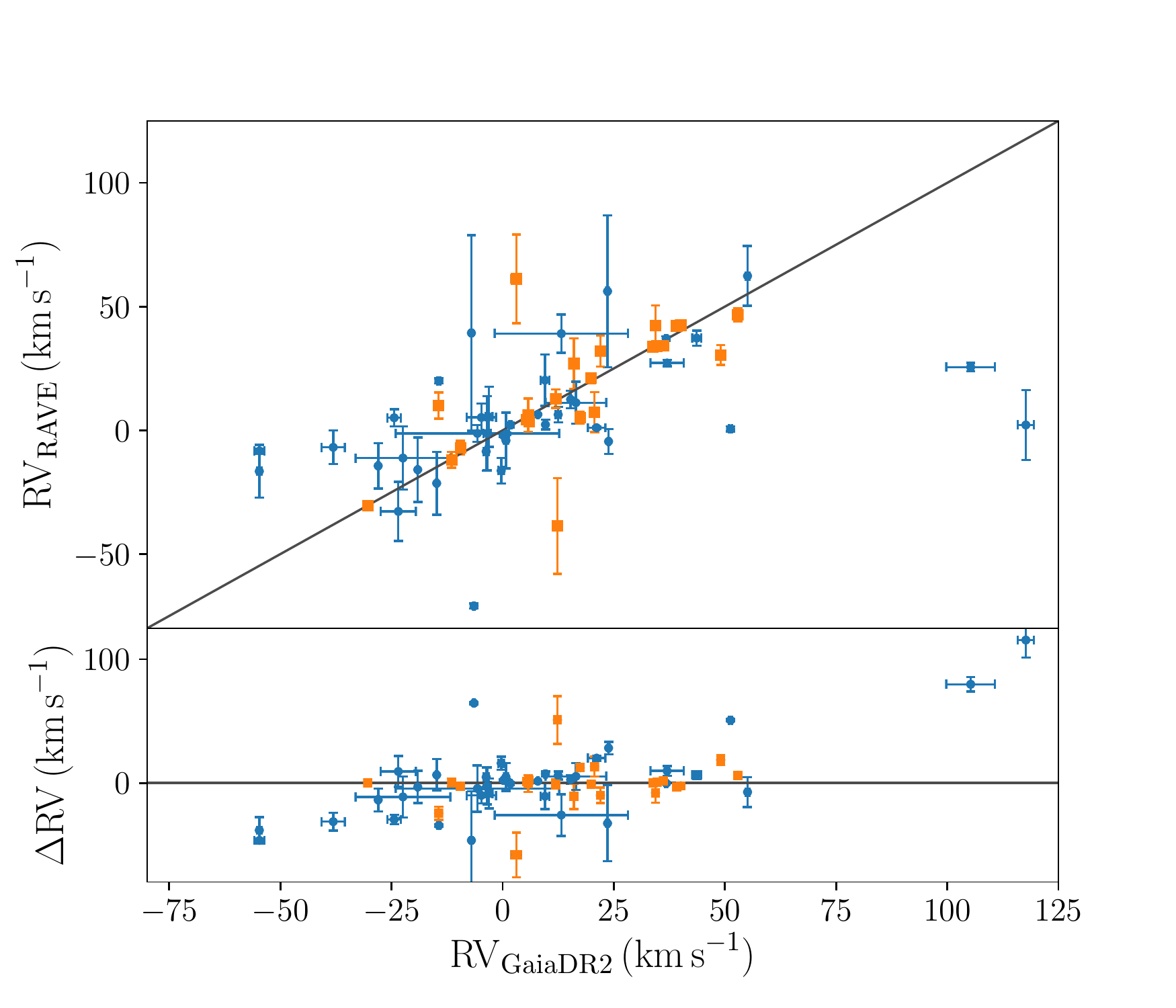}      
\includegraphics[width=\columnwidth]{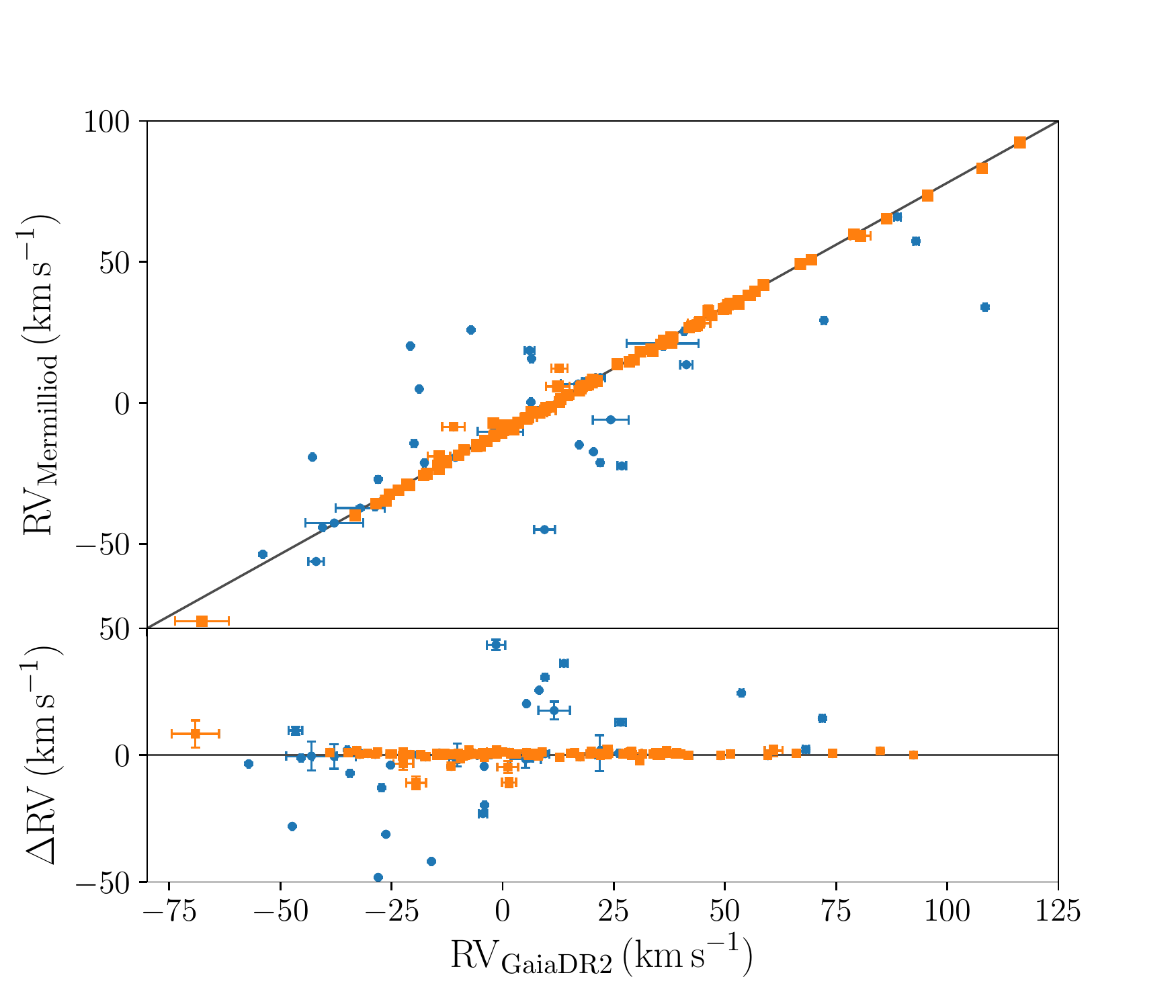}      
\caption{Comparison of mean RV from \gdrtwo\  with those from \cite{2014A&A...562A..54C} using RAVE data  (upper panel) and from \cite{mer08,mer09} (lower panel). The OCs with at least three members in each catalogue are shown in orange. The error bars combine the uncertainties of both catalogues.}
\label{f:RAVE_DR2}
\end{center}
\end{figure}

We compared our catalogue to that of \cite{2014A&A...562A..54C}, who provide the mean RV of 110 OCs based on RAVE DR4 \citep{2013AJ....146..134K}, 62 of which are common with our sample. The median difference for these common OCs is 0.1 \kms\ with a large dispersion (MAD= 8 \kms) because of several outliers with significant disagreement. The largest disagreement, 115 \kms, is for IC 2581, which has only one member in each catalogue. If the comparison is restricted to the 25 OCs with at least three members in each catalogue, the MAD decreases to 3 \kms, but still has several outliers, as shown in Fig. \ref{f:RAVE_DR2}. Several OCs have very large error bars in RAVE, such as IC 4729, NGC 2451A, and Alessi 24, which do not correspond to the RV error of individual stars. It is likely that non-members or spectroscopic binaries were included in the mean RV from RAVE for these OCs. On that point, the new astrometric memberships derived in Paper I  dramatically improve the reliability of the mean RVs.

Although it is included in the MWSC, we focus here on the homogeneous sample from \cite{mer08,mer09}, which provides mean RV for 172 OCs, 142 of which are part of our list. The comparison shows an excellent agreement (see Fig. \ref{f:RAVE_DR2}) with a median difference and MAD of 0.5 \kms. Twenty OCs show a difference of 10 \kms\ or larger. Most of these outliers (18) correspond to OCs with only one RV member, either in \gdrtwo\ or in  \cite{mer08}, so the significant difference may not be reliable. The two remaining targets, NGC 2925 and Trumpler 3, have less significant differences of $\sim$10 \kms\ based on four and five members from \gaia, and two and three stars in  \cite{mer08}. An interesting case is Stock 2 , which has a \gaia\ mean RV of $8.2\pm 0.1$ \kms\ based on 183 members, while the \cite{mer08} determination is based on one star ($-22.4\pm0.15$ \kms). The value from the MWSC does not agree for this OC, either, although it is based on 27 stars ($1.8\pm1.8$ \kms). \gdrtwo\ provides a dramatical improvement of the mean RV of this OC. 

The OCCASO survey focuses on red giants in OCs that were poorly studied before. It observes them at high spectroscopic resolution in order to derive RVs and abundances \citet{2016MNRAS.458.3150C}. Eighteen OCs have been observed with at least six giants per cluster, and mean RVs were determined with a median precision of 0.2 \kms\ \citep{2017MNRAS.470.4363C,Laia_OCCASO}. The 18 OCs are part of our catalogue and we derived a median difference less than 0.1 \kms\ with a MAD of 0.3 \kms. Only 2 OCs differ by more than 0.6 \kms\ (but less than 1.6 \kms), NGC 6705 and NGC 6791, which have standard deviations of $\sim$1.6 \kms\ in both catalogues. 

These comparisons show the excellent agreement of the \gdrtwo\ RVs with ground-based catalogues, as has been reported by \cite{2018arXiv180409372K}, and this confirms the good quality of the RVS data. The agreement with the homogeneous high-quality catalogues of \cite{mer08,mer09} and OCCASO \citep{2017MNRAS.470.4363C} is better than 0.5 \kms. 

Most of the outliers in these comparisons are well explained by the lowest reliability of the mean RVs that rely on very few members. This convinced us to select a high-quality subsample of OCs that include at least three RV members, giving an uncertainty on the mean RV lower than 3 \kms. This subsample of 406 OCs is used in the following to better interpret the OC kinematics.

\subsection{High-quality sample}

For the high-quality sample, the median number of members is seven and the median uncertainty of the weighted mean RV is 0.5 \kms.  \gdrtwo\  considerably improves the kinematical information of these 406 OCs, both in the number of members with a RV determination and in the precision of the mean value for the OC. \gdrtwo\ provides a first RV for several OCs. For some OCs that have an RV available in the literature,  many more members are provided on which the mean RV can be determined. We list below a selection of the most remarkable improvements due to \gdrtwo\ RV data.

\begin{itemize}
\item Of the 60 newly discovered OCs, named Gulliver and reported in Paper I, 11 are part of the high-quality sample with up to 12 members (Gulliver 9) and 15 members (Gulliver 6). Gulliver 6 is also the nearest of these newly discovered OCs, at 416 pc. Gulliver 4 and Gulliver 44 show the lowest RV dispersion, $\sigma_{\rm RV}=1$ \kms, with three and five members, respectively.
\item We report 49 RV members for Collinder 110, which is part of the MWSC, but has no RV provided there. The recent determination for Collinder 110 of $38.7\pm0.8$ \kms\ by \cite{2014AJ....147..138C} agrees excellently well with our determination of $38.2\pm0.2$ \kms. 
\item We report 51 RV members for Roslund 6, which is also part of the MWSC, but lacks an RV there. Roslund 6 has no RV in the literature as yet, although it is a nearby OC at 352 pc.
\item We report the first RV determination for Pismis 3: RV$=30.3\pm0.25$ \kms\ . This is based on 33 members.
\item We report the first RV determination for Stock 1: RV$=-19.5\pm0.5$ \kms\ . This is based on 30 members.
\item Pozzo 1 corresponds to the $\gamma$ Velorum cluster, or Vela OB2, whose complex kinematical structure has been studied by \cite{2014A&A...563A..94J}  as part of the \gaia\ ESO Survey \citep{2012Msngr.147...25G, 2013Msngr.154...47R}. \cite{2014A&A...563A..94J} found two groups separated by 2 \kms. We find 27 members in that area with a mean value of RV$=18.7\pm0.7$ \kms\ , corresponding to a clear peak, while a secondary peak might lie at RV=16.8 \kms. This agrees well with \cite{2014A&A...563A..94J}. However, the RV uncertainties for these stars are large, with a median value of 4.4 \kms. This means that these stars are fast rotators, as has also been shown by \cite{2014A&A...563A..94J}.
\item Trumpler 19 is a poorly studied OC for which we report 26 RV members and RV$=-26.4\pm0.3$ \kms.
\item ASCC 41, or Herschel 1, is a poorly populated OC confirmed by \cite{2011A&A...530A..32B} using photometry. We report the first RV determination for it: RV$=-10.1\pm0.35$ \kms\ . This is based on 23 members.
\item Ruprecht 147 is the oldest nearby star OC, at a distance of $\sim$300 pc with an age of $\sim$3 Gyr. It received little attention until the spectroscopic study by \cite{2013AJ....145..134C}, which gave an RV of 41.1 \kms\ that agrees very well with our determination of RV$=41.8\pm0.15$ \kms\ , which is based on 72 members. The value determined by \cite{2014AJ....147..138C}, RV$=42.5\pm1.0$ \kms, also agrees well.
\item The Pleiades have the most RV members, with 212 members giving RV$=5.9\pm0.1$ \kms. This doubles the number of RV members quoted in MWSC. 
\item At a distance similar to that of the Pleiades ($\sim$135 pc), Platais 8 benefits from a significant improvement of its mean RV thanks to 32 members, which give RV$=20.6\pm0.6$ \kms.
\item The second most populated OC is NGC 3532 with 209 members and RV$=5.4\pm0.2$ \kms. In the MWSC, its RV was based on four stars.
\item The nearest OC in our high-quality sample is Alessi 13 (104 pc), which has three members and an RV$=21.1\pm2.0$ \kms.
\end{itemize}

About half of our high-quality sample had either no RV in the literature or an RV based on only one member in the MWSC.

\begin{figure}[ht!]
\begin{center}
\includegraphics[width=\columnwidth]{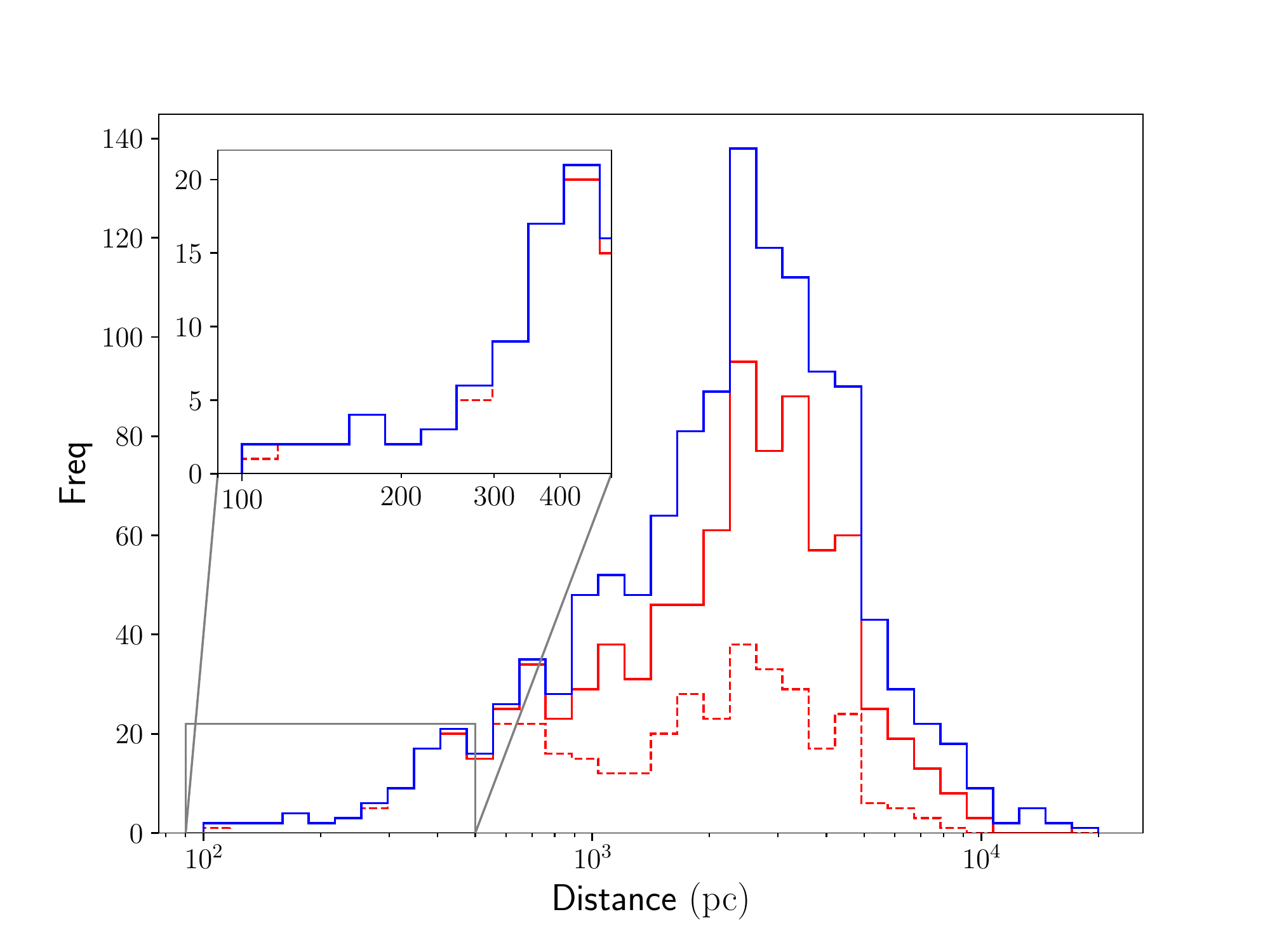}      
\caption{Histogram of OC distances as listed in Paper I (in blue) for the 861 OCs with a \gdrtwo\ RV determination (in red) and for the 406 OCs of the high-quality sample (dotted red). A zoom on nearby OCs is presented in the inset panel.}
\label{f:histo_dist}
\end{center}
\end{figure}

Figure \ref{f:histo_dist} shows the histogram of distances based on \gdrtwo\ astrometry determined in Paper I for the full catalogue of 1\,237 objects as well as the 861 OCs with an RV determination, and the 406 OCs of the high-quality sample. The zoom on the nearby OCs shows that our high-quality sample is complete at 99\% up to 500 pc, compared to the Paper I list. As explained in Paper I, three nearby OCs are not part of our sample because of their large extension on the sky, namely  Collinder 285 (the Ursa Major moving group), Melotte 25 (the Hyades), and Melotte 111 (Coma Ber).  The parameters of these nearby OCs are provided in \cite{2018A&A...616A..10G}. All the other OCs from Paper I closer than 500 pc to the Sun are part of our high-quality sample, except for 3 OCs. Mamajek 1 (104 pc) is missing because it has only two RV members, which give a mean of RV$=16.7\pm0.9$ \kms. NGC 1333 (296.5 pc) is missing because it has only one member with a large uncertainty at RV$=1.9\pm5.9$ \kms. The dark cloud nebula LDN 1641 South  at 432 pc has no RV measurement. It is worth noting that nearby OCs were discovered thanks to \gdrtwo\ by \cite{2018arXiv180503045C} after the list in Paper I was established.

Compared to the MWSC, several nearby OCs seem to be missing in our high-quality sample. Some OCs that are listed in the MWSC with a distance closer than 500 pc that we did not find  in Paper I may be explained mainly by the low contrast of members with respect to the background, in density, and in proper motion. Loose associations, such as the $\epsilon$ Chamaeleontis and the $\mu$ Oph groups, may have been missed for this reason, as explained in Paper I. In other cases, the distance estimation in MWSC is in question. For instance, Collinder 132 has an estimated distance of 330 pc in the MWSC, but it is 653.5 pc in Paper I based on the parallax of nearly 100 stars. The same situation occurs for Stock 23: 450 pc in the MWSC and 609 pc in Paper I. The MWSC also contains several objects that were shown to be not real  \citep[see e.g.][]{2018MNRAS.480.5242K}. Even if it is not complete for the associations and moving groups, our high-quality sample is fairly complete in terms of OCs up to 500 pc, which gives us the opportunity to investigate their kinematics. 

\section{Galactic velocities}
\label{s:velocities}
We computed heliocentric and Galactic Cartesian and cylindrical positions and velocities of the 861 OCs by combining their mean position, most probable distance, proper motion, and associated uncertainties from Paper I with their weighted mean RV and uncertainty determined in this study. These positions and velocities are provided in an electronic table at the CDS. We adopted the same conventions and reference values for the Sun as in \cite{2018A&A...616A..11G}. The Cartesian U, V, and W velocities with respect to the Sun are oriented towards the Galactic centre, the direction of Galactic rotation, and the north Galactic pole, respectively. The Cartesian Galactic coordinates (X, Y, Z) are such that the Sun is at X=-8.34 kpc, Y=0, Z=27 pc. The Galactic cylindrical coordinates are 
$(R, \phi, Z, V_{\rm R}, V_\phi, V_{\rm Z})$ 
with 
$\phi$ in the direction of Galactic rotation and the origin at the line Sun-Galactic centre. The circular velocity at the solar radius is $V_c = 240$ \kms, and  the peculiar velocity of the Sun with respect to the local standard of rest is (U, V, W) = (11.1, 12.24, 7.25) \kms. With the error propagation, we obtain median velocity uncertainties of (0.6, 0.8, 0.2) \kms\ in $(V_{\rm R}, V_\phi, V_{\rm Z})$ and (0.4, 0.5, 0.1) \kms\ when the high-quality sample is considered. Only two OCs have an uncertainty in one velocity component larger than 20 \kms, BH 222 and Berkeley 29. BH 222 is a starburst cluster in the inner Milky Way according to \cite{2014A&A...567A..73M} for which our RV determination is reliable (RV$=-119.3\pm2.8$ \kms\ based on five stars) and in agreement with that of \cite{2014A&A...567A..73M}. However, the astrometric parameters are more uncertain, which leads to velocity uncertainties of  $\sim$100 \kms\ in the V$_{\rm R}$ and V$_\phi$ components. For Berkeley 29, we have only one candidate member, with a membership probability $p=0.4$, which differs by 25 \kms\ from the mean value of the cluster determined by \cite{2005AJ....130..597Y}. This OC is the most distant open cluster known, at a Galactocentric distance $>$20 kpc  \citep[e.g. ][]{2004MNRAS.354..225T}. The brightest stars of that cluster have  $V > 14.2$  and were confirmed as RV members by \cite{2005A&A...429..881B}, who determined RV$\sim 29$ \kms\ with low-resolution spectra. Another determination gives RV$=24.66 \pm 0.36$ \kms\ \citep{2008A&A...488..943S}. In conclusion, the only star on which our RV is based is likely not a member of Berkeley 29, and the mean value given in our catalogue is incorrect (a note is provided in the online table). Except for these two objects, the unprecedented precision in 3D velocities of our sample allows us to revisit the kinematics of the OC population. 

\subsection{Nearby open clusters}
\begin{figure}[ht!]
\begin{center}
\includegraphics[width=\columnwidth]{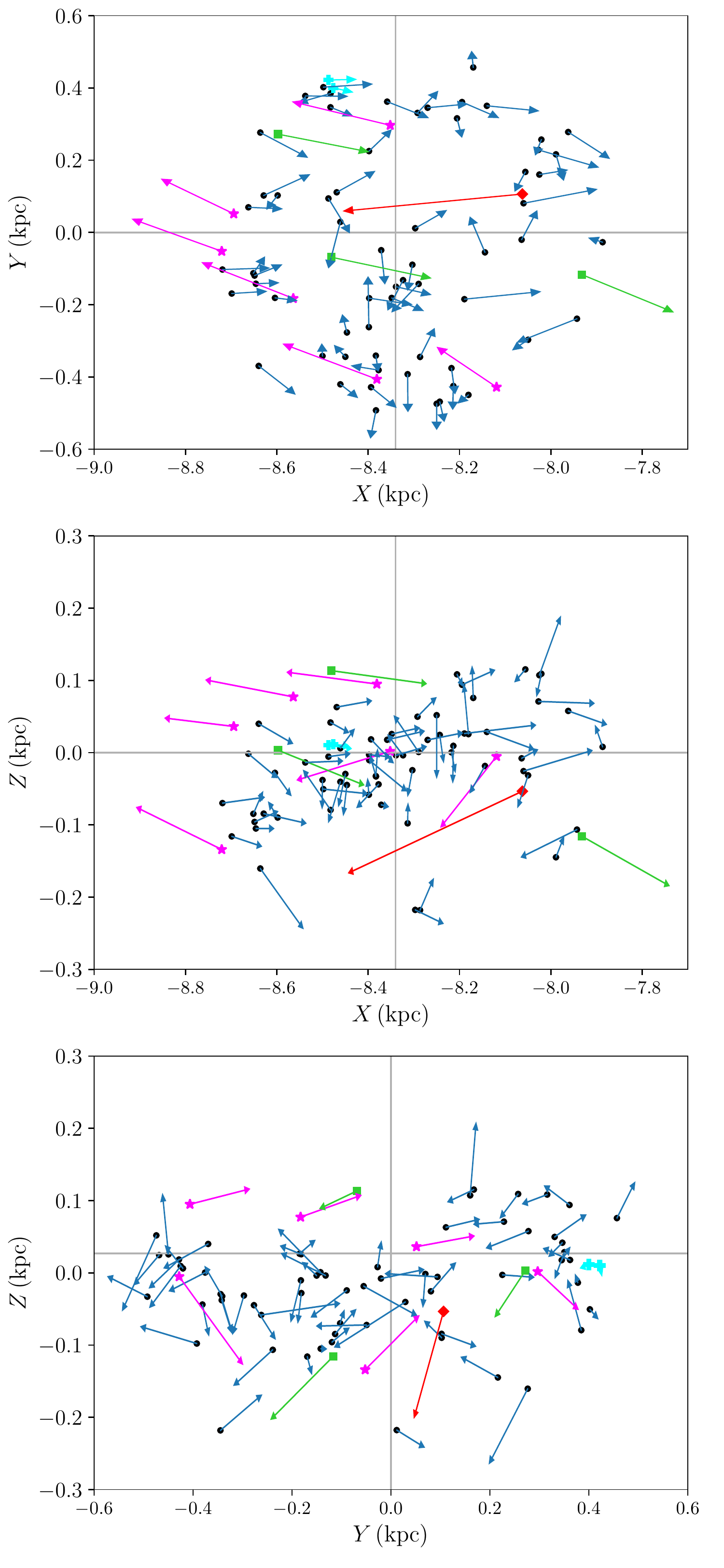}      
\caption{Galactic position in (X, Y) (upper panel) and (X, Z) (middle panel) and (Y, Z) (bottom panel) for the high-quality OC sample closer than 500 pc. The size and angle of the arrows are proportional to $(V_{\rm R}, V_\phi - V_c), (V_{\rm R}, V_{\rm Z}),$ and $ (V_\phi - V_c,  V_{\rm Z})$ . The horizontal and vertical lines indicate the position of the Sun. Several objects discussed in the text are represented by different symbols and colours. The red diamond represents Ruprecht 147, the green squares show Mamajek 4, NGC 2632, and Stock 2 (Group 1 in Table \ref{t:groups_age}), the magenta stars show Turner 5, NGC 7092, ASCC 41, NGC 1662, Ruprecht 98, and Stock 10 (Group 2 in Table \ref{t:groups_age}), and the blue circles indicate RSG 7 and RSG 8. }
\label{f:XYZ_near}
\end{center}
\end{figure}

Figure \ref{f:XYZ_near}  displays the distribution of the nearby OCs (distance $\le$ 500 pc) of the high-quality sample in position with a representation of their motion. The visual inspection of these plots provides the following information:

\begin{itemize}
\item The OC with the highest velocity is Ruprecht 147, which is expected since it is the oldest cluster in the solar neighbourhood (log(age)\footnote{log(age in yr)}=9.33 in the MWSC) and thus its kinematics is representative of the hot thin disc.
\item In the OCs with the largest motion, two groups can be seen that share similar velocities in the three components. The first group includes Mamajek 4, NGC 2632, and Stock 2. They are characterised by a significant radial motion towards the Galactic centre, a small rotation lag, and a negative vertical motion. These OCs are not close in space. The second group includes Turner 5, NGC 7092, ASCC 41, and NGC 1662, and possibly also Ruprecht 98 and Stock 10, all characterised by  a significant radial motion towards the Galactic anticentre with a rotation slightly faster than that of the LSR. The 6D phase-space parameters of the OCs in these two groups are detailed in Table \ref{t:groups_age}.
\item RSG 7 and RSG 8, two OCs recently found by \cite{2016A&A...595A..22R}, are confirmed as two separate OCs very close in space and motion.
\item The cluster Turner 5 has a velocity that is closest to that of the Sun. According to the MWSC, it is much younger and more metal-poor than the Sun, which excludes this OC as the birthplace of the Sun. Two other OCs differ by less than 10 \kms\ from the velocity of the Sun, RSG 5 and Teutsch 35. RSG 5 is a very young OC according to \cite{2016A&A...595A..22R}, while there is no information yet on the properties of Teutsch 35.
\end{itemize}

\begin{figure}[ht!]
\begin{center}
\includegraphics[width=\columnwidth]{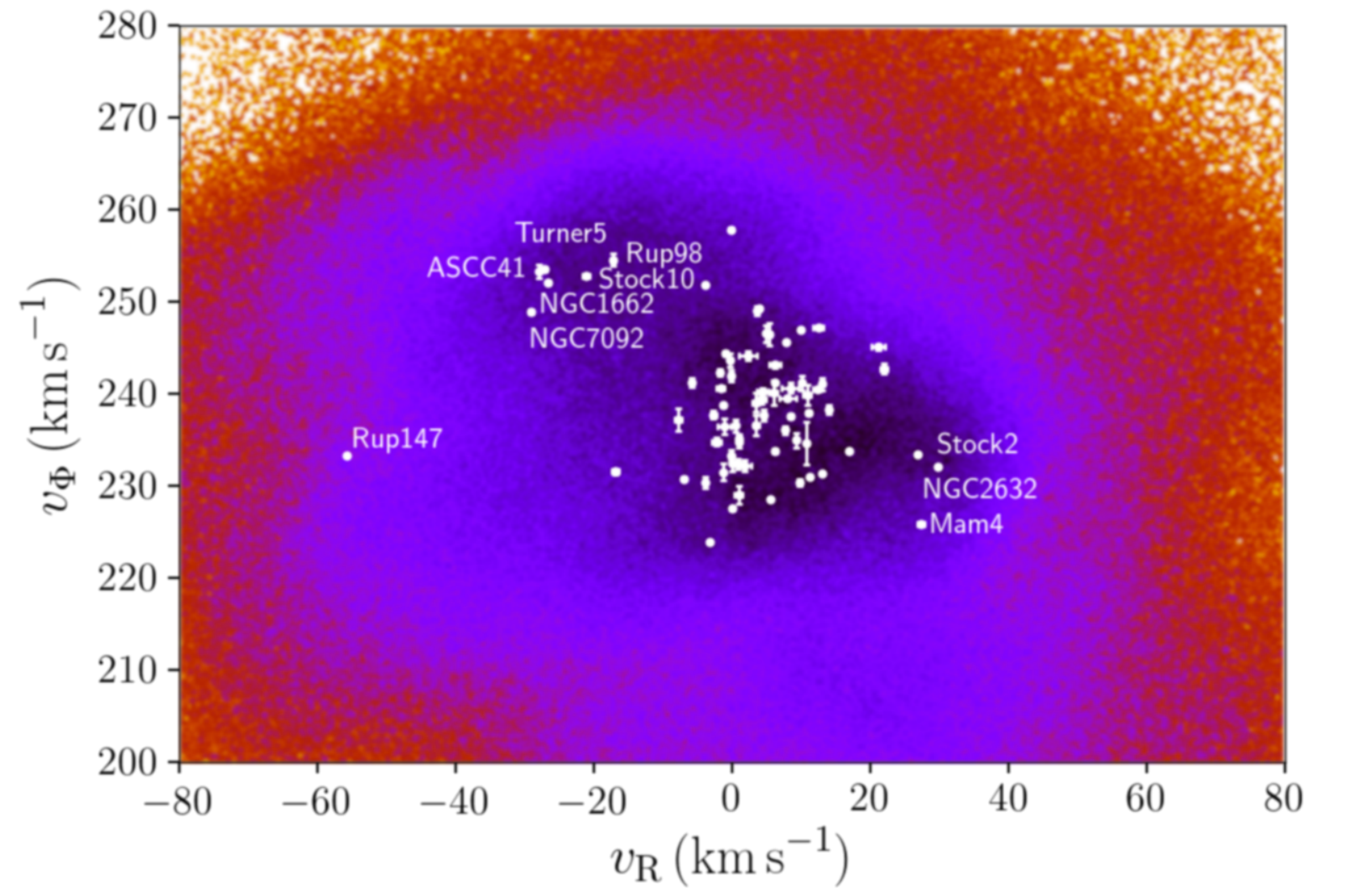}      
\caption{Velocity distribution of nearby OCs (dist $\le$ 500 pc) in $(V_{\rm R}, V_\phi)$, superimposed on field stars in the same volume around the Sun taken from \cite{2018A&A...616A..11G}. The most extreme OCs are indicated, in particular, the two groups listed in Table \ref{t:groups_age}.}
\label{f:VrVphi_near}
\end{center}
\end{figure}

How do the kinematics of the nearby OCs compare with the kinematics of field stars? It is known from the Hipparcos era \citep{1997A&A...323L..49P} that the stellar phase-space distribution in the solar neighbouhood is clumpy \citep[e.g.][]{1998MNRAS.298..387D, 1999A&AS..135....5C, 2005A&A...430..165F, 2008A&A...490..135A}. This has been confirmed by \gdrtwo\ data with a high degree of detail by \cite{2018A&A...616A..11G}, who showed that several nearby OCs are associated with overdensities in the (U,V) plane. Seven of the eight OCs that lie closer than 200 pc  were shown to be associated with a  large structure that forms an arch. We show in Fig. \ref{f:VrVphi_near} the $(V_R, V_\phi)$ distribution of our OC sample that lies closer than 500 pc together with the stellar distribution in the same volume, taken from the catalogue built by \cite{2018A&A...616A..11G}. The OCs clearly overlap with the field star clumps that have the highest density. Only Ruprecht 147 is isolated: its velocity is significantly different from that of the other OCs, but is still compatible with the field star velocity. The two groups that are identified in Fig. \ref{f:XYZ_near} based on their higher velocity, in particular in the radial direction, stand out in  Fig. \ref{f:VrVphi_near}, where they lie on diametrically opposed sides of the bulk distribution. However, they are clearly associated with the field star structures. Ages are available in MWSC for these OCs and reported in Table \ref{t:groups_age}. They are older than the majority of nearby OCs. The median log(age) of nearby OCs is $\sim$8.2 according to MWSC whereas these nine OCs with large velocities range from log(age)=8.42 to log(age)=8.92. Although error bars are not available for all of them, the ages of these OCs are rather different, which means that they are probably not physical groups but merely confirm the known correlation between age and kinematics in the thin disc.

\begin{table*}[h]
  \centering 
  \small
  \setlength\tabcolsep{2pt}
 \caption{ log(age) from the MWSC  and phase-space information from \gaia\  (in pc and \kms) for OCs  in two kinematical groups with higher radial motion (see text for the adopted conventions).  }
  \label{t:groups_age}
\begin{tabular}{llcccccc}
\hline
  cluster        & log(age) & X & Y & Z &  $V_{\rm R}$ & $V_\phi$ & $V_{\rm Z}$ \\  
\hline
 \multicolumn{8}{c}{Group 1} \\
\hline
Mamajek 4   & 8.815 &  -7931.9$\pm$1.1 &-116.8$\pm$0.3 & -115.7$\pm$0.3  &  27.3$\pm$0.5 & 225.8$\pm$0.2 & -9.8$\pm$0.2\\  
NGC 2632    & 8.92  &  -8480.8$\pm$0.1 & -68.5$\pm$0.1 &  113.5$\pm$0.1  &  29.8$\pm$0.1 & 232.1$\pm$0.1 & -2.6$\pm$0.1\\ 
Stock 2     & 8.44  &  -8597.2$\pm$0.2 & 272.0$\pm$0.2 &    3.3$\pm$0.1  &  26.8$\pm$0.1 & 233.4$\pm$0.1 & -6.9$\pm$0.1\\ 
 \hline
\multicolumn{8}{c}{Group 2} \\
\hline
Turner 5    & 8.49  &  -8380.9$\pm$0.3 &-406.3$\pm$3.2 &   94.9$\pm$0.6  & -28.0$\pm$0.2 & 253.3$\pm$0.8 &  2.3$\pm$0.2\\  
NGC 7092    & 8.569 &  -8351.6$\pm$0.1 & 296.6$\pm$0.7 &    1.8$\pm$0.1  & -29.1$\pm$0.1 & 248.9$\pm$0.2 & -5.7$\pm$0.1\\ 
ASCC 41     & 8.7   &  -8563.9$\pm$0.7 &-182.7$\pm$0.6 &   77.2$\pm$0.2  & -27.2$\pm$0.3 & 253.5$\pm$0.2 &  3.3$\pm$0.1\\ 
NGC 1662    & 8.695 &  -8720.6$\pm$1.1 & -52.1$\pm$0.2 & -134.1$\pm$0.4  & -26.6$\pm$0.2 & 252.1$\pm$0.1 &  8.3$\pm$0.1\\
Ruprecht 98 & 8.8   &  -8118.8$\pm$0.7 &-428.1$\pm$1.4 &   -5.2$\pm$0.1  & -17.2$\pm$0.3 & 254.6$\pm$0.7 &-13.8$\pm$0.1\\
Stock 10    & 8.42  &  -8694.3$\pm$1.0 & 51.6 $\pm$0.1 &   36.2$\pm$0.1  & -21.1$\pm$0.5 & 252.8$\pm$0.1 &  1.6$\pm$0.1\\ 
 \hline
\end{tabular}
\end{table*}

\subsection{Age velocity relation for open clusters}
\begin{figure}[ht!]
\begin{center}
\includegraphics[width=\columnwidth]{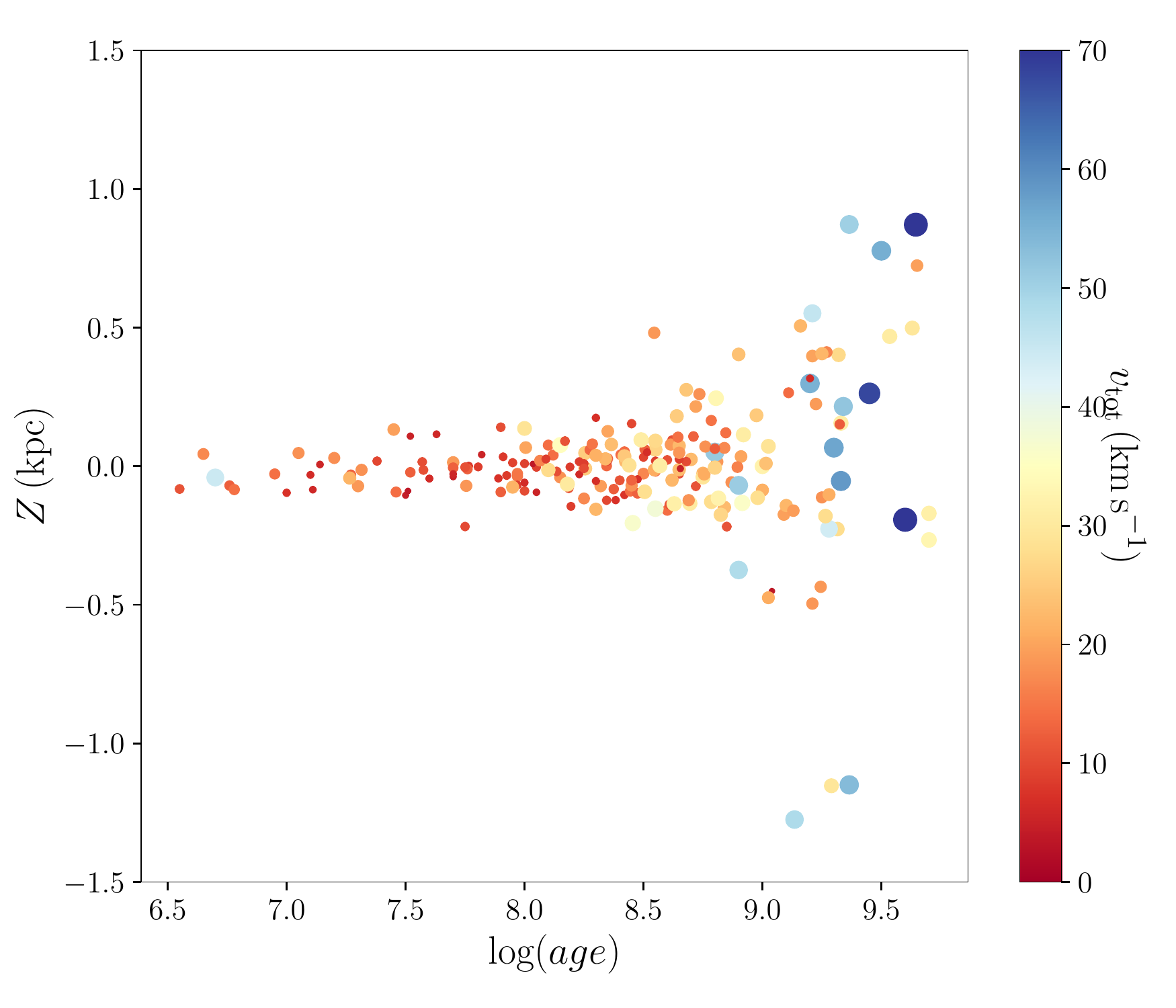}      
\caption{Distribution of the high-quality sample in Z, distance to the Galactic plane, vs. age. The size of the symbols is proportional to the total heliocentric velocity of the targets.}
\label{f:Z_age}
\end{center}
\end{figure}

We used the full sample of 861 OCs where mean velocities and dispersions were computed in four age groups. Ages are available in the MWSC for about half of the sample. A velocity ellipsoid was fitted in each age bin using the stochastic expectation maximization
(SEM) method \citep{SEM}, which separates multivariate Gaussian populations without any a priori information. This method has previously been used in stellar kinematics by \cite{2003A&A...398..141S} and \cite{2010A&A...516A...3K}. The results are presented in Table \ref{t:ellipsoid}. In each age bin, two populations were assumed, but as expected, the algorithm converged to a mixture where the bulk OC population dominates at more than 90\%, the rest represent the few OCs with different velocities. Although a velocity ellispoid is not a perfect representation in a non-axisymetric disc, it offers a convenient way to show how the kinematical behaviour of the OC population evolves with time. The most striking result is the low dispersion obtained for the youngest populations, in particular in V$_{\rm Z}$ , where the dispersion is 4-5 \kms. This has to be compared to the dispersions obtained by \cite{2009MNRAS.399.2146W} from their sample of $\sim$500 OCs extracted from DAML: $(\sigma U, \sigma V, \sigma W) = (28.7, 15.8, 11.0)$ \kms.  The dispersion in the three components increases smoothly with age with a significant increase for the oldest population.  
The oldest population is clearly very distinct from the younger ones, as is shown by the vertical distribution of the high-quality sample as a function of age in Fig. \ref{f:Z_age}. Where the young OCs are confined in the plane, the OCs with log(age) $\ge$ 9 exhibit a much wider range of vertical position, and their total velocity is higher as well. Moreover, the OCs with high velocities are old on average. The only OC that appears  with a high velocity and young age is NGC 2244, which was previously mentioned as a possible confusion with another OC in the Rosetta nebula (see Fig. \ref{f:NGC_2244}).

\begin{table}[h]
  \centering 
  \tiny
  \setlength\tabcolsep{2pt}
  \caption{Parameters of the velocity ellipsoid in four age bins. N is the number of OCs per bin. The values are those of the dominating population in a two-population fit, and the corresponding percentage of OCs is also given.}
  \label{t:ellipsoid}
\begin{tabular}{lrrrrrrrr}
\hline
  log(age) & $N$ &  \% &  V$_{\rm R}$&$\sigma_{\rm V_R}$&V$_\phi$& $\sigma_{V_\phi}$&V$_{\rm Z}$&$\sigma_{V_Z}$\\
yr & & & \kms\ & \kms\ & \kms\ & \kms\ & \kms\ & \kms\ \\
  \hline
 $<$ 7.8     & 106 & 90  &  -3.1   & 11.7  & 238.4   & 10.6   & -1.5   &  4.3\\
 7.8 -- 8.4  & 98 & 96  &  -1.0   & 13.2  & 238.6   & 11.1   & -0.0   &  5.2\\
 8.4 -- 9.0  &  109  & 95 &   0.8   & 17.1  & 237.7   & 11.8  &  -0.1   &  6.8\\
 $ \ge$ 9.0 & 61 & 93  &   6.0   & 22.3  & 233.8   & 23.2   &  1.0   & 14.0\\
  \hline
\end{tabular}
\end{table}

\subsection{Peculiar open clusters, pairs, and groups}
\begin{figure}[ht!]
\begin{center}
\includegraphics[width=\columnwidth]{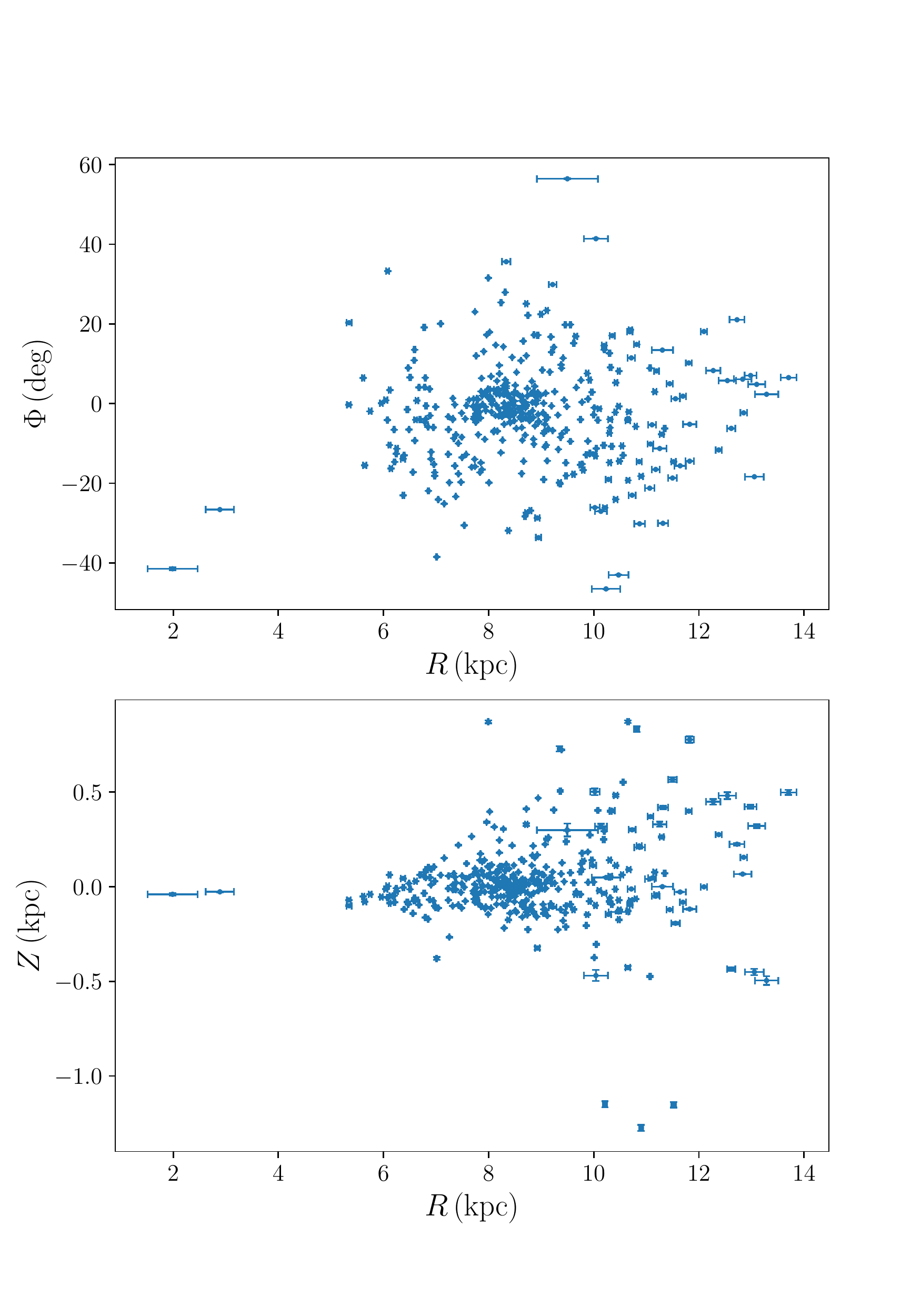}      
\caption{Spatial distribution of the high-quality sample in Galactic cylindrical coordinates.}
\label{f:RphiZ}
\end{center}
\end{figure}

Here we consider the high-quality sample of 406 OCs in order to identify OCs with unusual velocities, pairs of OCs, and groups. This sample spans distances up to $\sim$10 kpc, but most of the targets lie closer than 5 kpc. Figure \ref{f:RphiZ} shows the spatial distribution of  this sample in Galactic cylindrical coordinates. A few OCs are extreme in their Galactic position. The two inner OCs are BH 222 and Teutsch 85. BH 222 was mentioned previously as having a large error in distance. Teutsch 85 is a poorly studied OC for which we provide a good-quality RV of $-119.6\pm1.15$ \kms\ based on five stars. Three OCs lie $\sim$1 kpc below the Galactic plane: Melotte 66, NGC 2243, and NGC 2204.
Melotte 66 is one of the four OCs that were identified by \cite{2010MNRAS.407.2109V} to possibly have an extragalactic origin. However, it has been extensively studied since that paper. Its metallicity is well established \citep[$-0.36\pm0.03$, ][]{2016A&A...585A.150N}, and its chemical composition and age \citep[$3.4\pm0.3$ Gyr, ][]{2014A&A...566A..39C} make it a typical object of the thin disc. NGC 2243 is even more metal-poor \citep[$-0.50\pm0.08$,][]{2016A&A...585A.150N} and NGC 2204 to a lesser extent \citep[$-0.24\pm0.08$,][]{2016A&A...585A.150N}. For these two objects, the detailed chemical composition has been demonstrated to be typical of the thin disc by \cite{2011AJ....141...58J}. The new velocities obtained with \gaia\ data do not change these conclusions.

\begin{figure}[ht!]
\begin{center}
\includegraphics[width=\columnwidth]{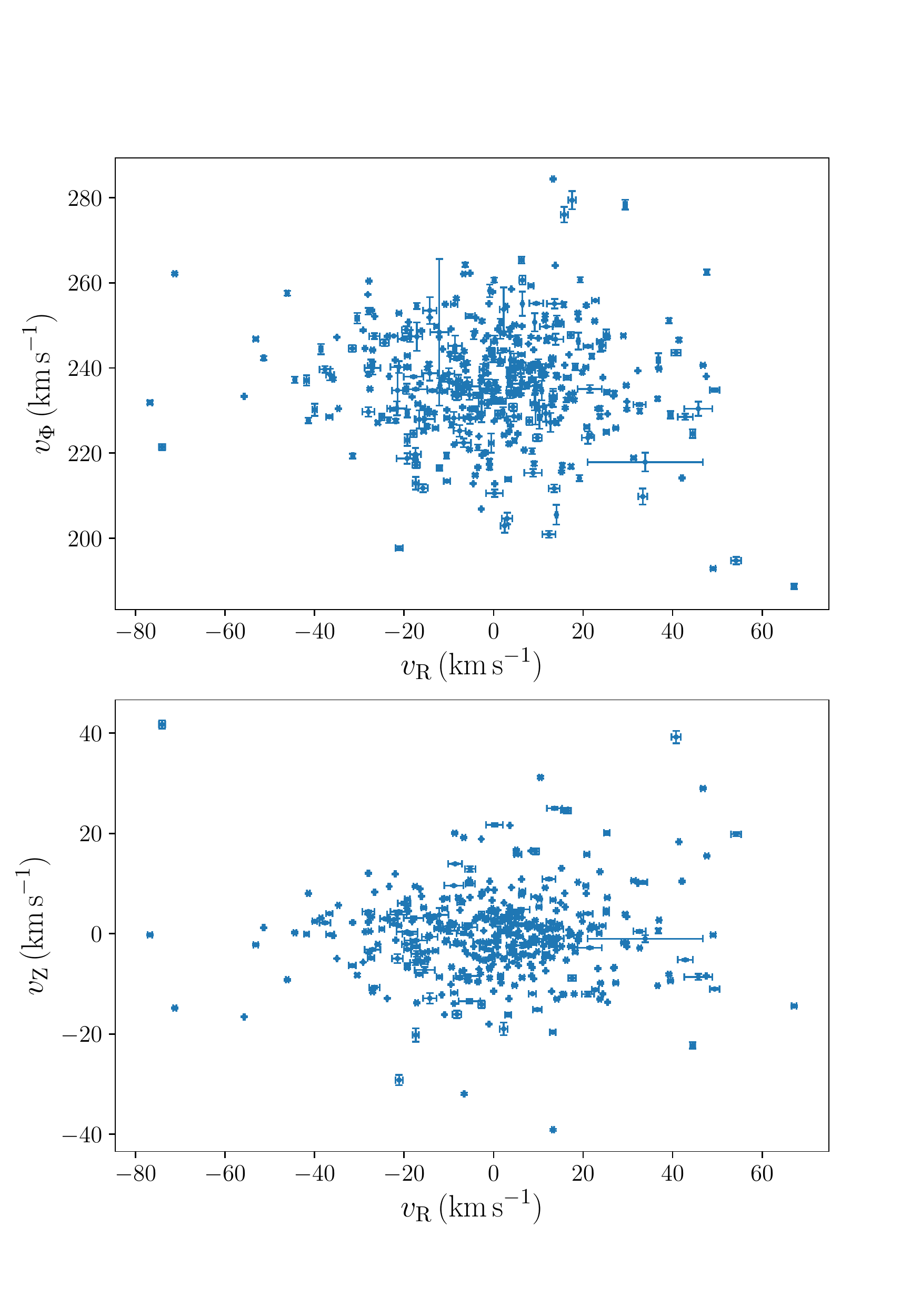}      
\caption{Velocity distribution of the high-quality sample in Galactic cylindrical coordinates. BH 222 and BH 140 are not represented (see text).}
\label{f:VRVphiVZ}
\end{center}
\end{figure}

Two clusters have unusually high velocities for OCs: BH 222 (already mentioned for its large parallax uncertainty) and BH 140.  BH 140 was poorly documented until Paper I, where it is shown to be a globular cluster.  Our RV determination based on four stars (RV$=90.4\pm0.9$ \kms) combined with the good astrometric parameters determined in Paper I from 439 members gives a 3D velocity typical of the halo and thus confirms the nature of this cluster. Figure \ref{f:VRVphiVZ} shows the velocity distribution of  the high-quality sample in Galactic cylindrical coordinates, without these two objects. 

Other OCs look extreme in their Galactic velocity. 
Trumpler 19, Haffner 5, and Berkeley 17 have $V_R < -70$ \kms,  BH 72, Ruprecht 171, FSR 1407, and Ruprecht 75 have $V_\phi > 270$ \kms,  and Berkeley 17, Berkeley 14, and Ruprecht 171 have $|V_Z| > 35$ \kms. Some of them (Trumpler 19, Haffner 5, Ruprecht 171, FSR 1407, and Ruprecht 75) have been poorly studied until now, but the MWSC lists them as old OCs, which are interesting for further study now that their members have been refined thanks to \gdrtwo.

\begin{figure}[ht!]
\begin{center}
\includegraphics[width=\columnwidth]{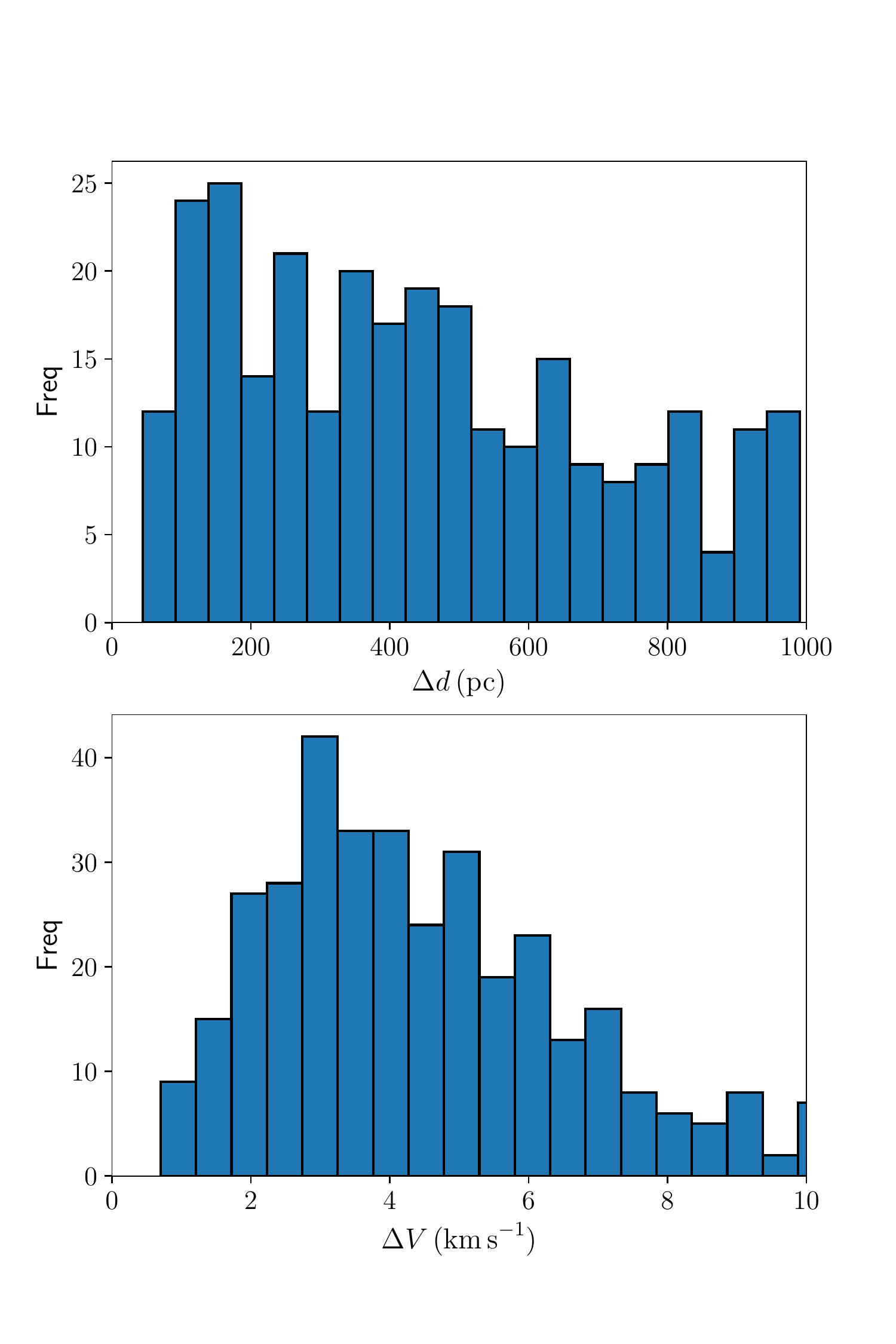}      
\caption{Histogram of the distance between nearest neighbours in the high-quality sample (upper panel) and the same in velocity (lower panel).}
\label{f:NN}
\end{center}
\end{figure}

Binary clusters are supposed to exist, but their fraction in the Galaxy, possibly from 8\% to 20\%, is subject to debate \citep[see e.g. ][ and references therein]{1995A&A...302...86S, 2009A&A...500L..13D}. Our high-quality sample gives a good opportunity to search for such objects. A pair of nearby clusters, RSG 7 and RSG 8, was found to be close in space and motion based on the examination of Fig. \ref{f:XYZ_near} (Sect. 3.1). With fewer than 100 nearby OCs, it was possible to see this pair in the figures. For the larger high-quality sample, it is mandatory to use more objective criteria that are able to measure the proximity of targets in the 6D phase space. \cite{2017A&A...600A.106C} adapted a method inspired by extragalactic work in order to identify OC groups with linking lengths 100 pc and 10 \kms. The linking length is related to the separation of members belonging to a group and should be significantly smaller than the typical separation of objects in the population. In order to estimate the typical separation of OCs in our sample, we looked for the nearest neighbour of each OC in distance and velocity. The corresponding histograms are shown in Fig. \ref{f:NN}. The separations in space peak at $\sim$150 pc, while the separations in velocities  peak at $\sim$3 \kms. This latter value is very low but not surprising owing to small dispersions that were found when fitting a velocity ellipsoid to our sample. Among the young population of OCs that dominates the sample, it is likely to find OCs that share very similar velocities. For instance, we find 39 cluster pairs with velocity differences lower than 2 \kms, although they do not show any peculiar proximity in space. The spatial separation is possibly more discriminant if we expect to find clusters that originated from a common molecular cloud in one possibly sequential star formation event. The closest pair of our sample includes ASCC 16 and ASCC 21, which are separated by $\sim$45 pc, with a velocity difference of 4.4 \kms. This is our best candidate for a physical pair.  Their physical link is supported by their similar young age, log(age) = 7. and 7.11, respectively, in the MWSC, and log(age) = 6.93 and 7.11 in DAML. Collinder 140  and   NGC 2451B form our second best candidate, separated by $\sim$ 58 pc, with a velocity difference of 1.9 \kms. This pair also has a similar young age, log(age) = 7.548 and 7.648 in DAML. A third pair with a separation smaller than 100 pc includes IC 2602 and Platais 8,  which are part of the binary OCs identified by  \cite{2017A&A...600A.106C}. The similarity of their age in DAML,  log(age) = 7.507 and 7.78, respectively, makes the physical binarity plausible.  The space separations for these three pairs have to be compared to the typical value of 10 pc of the binary candidates proposed by \cite{2009A&A...500L..13D}. Table \ref{t:pairs} gives the list of the pairs that differ by less than 200 pc in distance and 5 \kms\ in velocity in our high-quality sample. A possibly larger complex is formed by ASCC 16, ASCC 19, ASCC 21, Gulliver 6, and NGC 2232 since they appear several times in that table. RSG 7 and RSG 8 that were mentioned in the previous section are recovered with this method, but their separation in position is rather large. Precise ages and chemical composition would help to clarify the physical link of these pairs.

The most famous binary cluster is formed by $h$ and $\chi$ Persei \citep[NGC 869 and NGC 884, ][]{1913AAHam...2b...1M}. Unfortunately, these two OCs are not part of our high-quality sample because their mean RV relies on one and two stars, respectively. However, since they are part of the larger sample of 861 OCs, their separation in space and velocity can still be computed. They appear to be at $\sim$62 pc from each other. Similarly, we tested several  other binary candidates from the literature that could be retrieved in our full sample of 861 OCs even if the velocity difference is not reliable in some cases. The results are presented in Table \ref{t:binary_candidates}. None of the candidates proposed by \cite{2009A&A...500L..13D} are part of our sample, unfortunately. Several pairs identified by \cite{2017A&A...600A.106C}  were retrieved, among which Alessi 21 and  NGC 2422 are in the high-quality sample, with a reliable velocity difference, as well as  Platais 8  and IC 2602 mentioned above.   At $\sim$172 pc from each other, Alessi 21 and  NGC 2422 are more likely to result from a  chance alignment than that they are a physical binary system. An excellent candidate binary is the pair Collinder 394 and  NGC 6716, which lie at a distance of less than 30 pc from each other. They have different ages in the MWSC, but these ages are uncertain since no error bars are provided. 

 \begin{table}[h]
  \centering 
  \small
  \caption{Pairs of OCs differing by less than 200 pc in their Galactic position and 5 \kms\ in velocity in the high-quality sample.}
  \label{t:pairs}
\begin{tabular}{llrr}
\hline
 Cluster 1 & Cluster 2 & $\Delta$pos (pc) &  $\Delta$V (\kms) \\
 \hline
ASCC 101       &   NGC 7058        &    185   &  1.8 \\ 
 ASCC 105       &   Roslund 5       &    130   &  3.9 \\
 ASCC 16        &   ASCC 19         &    151   &  3.9 \\
 ASCC 16        &   ASCC 21         &     45   &  4.4 \\
 ASCC 19        &   Gulliver 6      &    181   &  4.3 \\
 ASCC 97        &   IC 4725         &    145   &  3.4 \\
 Alessi 20      &   Stock 12        &    183   &  2.7 \\
 Collinder 140  &   NGC 2451B       &     58   &  1.9 \\
 Gulliver 6     &   NGC 2232        &    159   &  4.8 \\
 IC 2602        &   Platais 8       &     83   &  4.5 \\
 RSG 7          &   RSG 8           &    145   &  2.8 \\
\hline
\end{tabular}
\end{table}

\begin{table}[h]
  \centering 
  \small
  \caption{Separation in space and velocity of binary candidates from the literature, computed from our mean parameters for 861 OCs.  Reference 1 is \cite{2017A&A...600A.106C}, reference 2 is \cite{1913AAHam...2b...1M}, reference 3 is \cite{2015MNRAS.453..106D}, and reference 4 is \cite{2016MNRAS.458.3150C}}
  \label{t:binary_candidates}
\begin{tabular}{lllrr}
\hline
 Cluster 1 & Cluster 2 & ref & $\Delta$pos (pc) &  $\Delta$V (\kms) \\
 \hline
 Alessi 13     &  Mamajek 1     & 1 &    292 &    5.5\\
 Alessi 21     &  NGC 2422      & 1 &    172 &     8.9\\
 Platais 8     &  IC 2602       & 1 &     83 &       4.5\\
 Turner 9      &  ASCC 110      & 1&   1759 &      6.0\\
 Collinder 394 &  NGC 6716      & 1 &     29 &     13.9\\
 IC 1396       &  NGC 7160      & 1 &    128 &     13.9\\
 NGC 869       &  NGC 884       & 2 &     62 &    19.7\\
 NGC 5617      &  Trumpler 22   & 3 &    559 &    10.8\\
 IC 4756       &  NGC 6633      & 4  &    375 &      6.7\\
 \hline
\end{tabular}
\end{table}




\section{Conclusions}
 \gdrtwo\ substantially improves our knowledge of the OC population. The astrometric membership from Paper I  cross-matched with the \gdrtwo\ RV catalogue of $\sim$7 million stars allowed us to compute the mean RV of  861 OCs. Particularly robust is our high-quality sample of 406 OCs for which the mean RV relies on at least three members. The median uncertainty of the mean RVs in this sample is 0.5 \kms. This list contains several poorly studied OCs that had no RV before, or a RV that relied on one or two questionable members. For the OCs that had been well studied before, the comparison with the best ground-based RVs shows a general agreement at the 0.5 \kms\ level. 
 
 These new RVs combined with the most probable distances and mean proper motions determined in Paper I allowed us to compute the 6D phase-space information of OCs. They were found to  follow the velocity distribution of field stars in the close solar neighbourhood that was previously revealed by \gdrtwo\ \citep{2018A&A...616A..11G, 2018arXiv180410196A}. As expected, the vertical distribution of young OCs is very flat, but the novelty is the high precision to which this can be seen. The dispersion of vertical velocities of young OCs is at the level of  5 \kms.  Clusters older than 1 Gyr span distances to the Galactic plane of up to 1 kpc with a vertical velocity dispersion of 14 \kms, typical of the thin disc. There is no need to invoke an extragalactic origin to explain the kinematical behaviour of the old OCs. 
 
 Five pairs of clusters with similar velocities were found with a separation of 29 to 83 pc, but none at a close separation of 10 pc as found in the Magellanic Clouds. This might be due to the incompleteness of our sample or to the low fraction of multiple systems forming in the local spiral arms. One group has five members that may be physically related. Other binary clusters previously identified have large separation in position and may correspond to a non-physical double, although further study with precise age and chemical composition is required to shed light on their nature. 
 
\begin{acknowledgements}
 This work has made use of results from the European Space Agency (ESA) space mission \gaia, the
data from which were processed by the \gaia\ Data Processing and Analysis Consortium (DPAC).
Funding for the DPAC has been provided by national institutions, in particular the institutions
participating in the \gaia\ Multilateral Agreement. C.S. and L.C. acknowledge support from the "programme national cosmologie et galaxies" (PNCG) of CNRS/INSU.
This work was supported by the MINECO (Spanish Ministry of Economy) through grant ESP2016-80079-C2-1-R (MINECO/FEDER, UE) and ESP2014-55996-C2-1-R (MINECO/FEDER, UE) and MDM-2014-0369 of ICCUB (Unidad de Excelencia 'Mar\' ia de Maeztu').
U.H. acknowledges support from the Swedish National Space Agency (SNSA/Rymdstyrelsen).
AB acknowledges PREMIALE 2015 MITiC. AKM acknowledges the support from the Portuguese Funda\c c\~ao para a Ci\^encia e a Tecnologia (FCT) through grants SFRH/BPD/74697/2010, and from the ESA contract AO/1-7836/14/NL/HB. AM and AKM acknowledge the support from the Portuguese Strategic Programme UID/FIS/00099/2013 for CENTRA.
The preparation of this work has made extensive use of Topcat \citep{2011ascl.soft01010T}, and of NASA's Astrophysics Data System Bibliographic Services. The \gaia\ mission website is
\url{http://www.cosmos.esa.int/gaia}.

\end{acknowledgements}

\bibliographystyle{aa}
\bibliography{OC_kinematics_V4}

\end{document}